\documentclass[acmtog]{acmart}
\acmSubmissionID{606}

\usepackage{booktabs} 

\usepackage{bm}
\usepackage[normalem]{ulem}
\usepackage[ruled]{algorithm2e} 

\SetAlFnt{\small}
\SetAlCapFnt{\small}
\SetAlCapNameFnt{\small}
\SetAlCapHSkip{0pt}

\acmJournal{TOG}

\begin{document}
\title{Anatomically Detailed Simulation of Human Torso}

\begin{teaserfigure}
\centering
\includegraphics[width=0.85\textwidth]{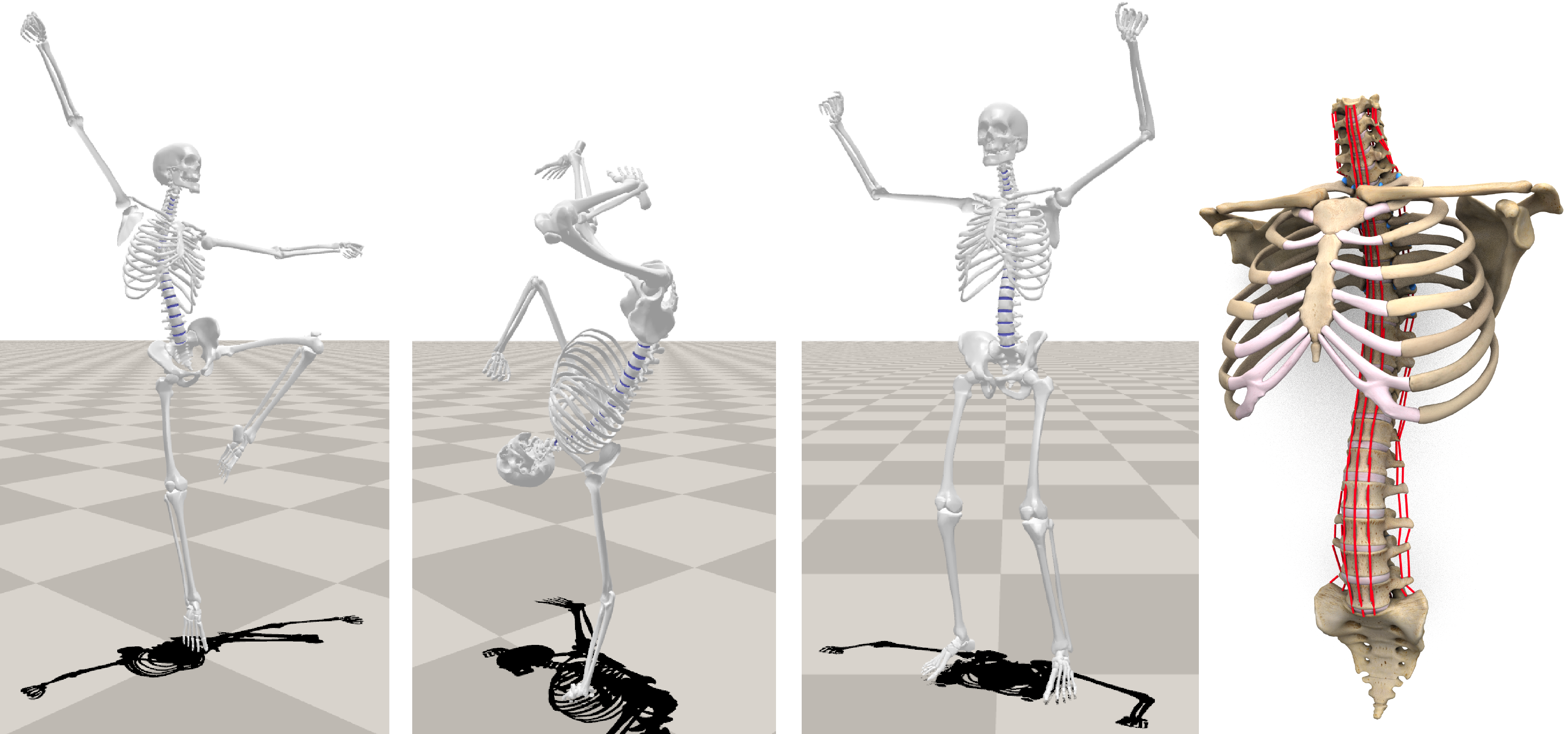}   
\caption{ (Left) Our method recovers underlying spine and shoulder bone movements unobservable by standard motion capture systems. (Right) Our torso model includes veterbrae bones, ligaments, facet joints and discs, scapula, clavicles, and a rib cage to generate physically and physiologically plausible movements.}
\label{fig:teaser}  
\end{teaserfigure}

\author{Seunghwan Lee}
\orcid{0009-0005-0892-2207}
\affiliation{%
 \institution{Stanford University}
 \streetaddress{450 Serra Mall}
 \city{Stanford}
 \state{CA}
 \postcode{94305}
 \country{USA}}
\email{lsw9021@gmail.com}

\author{Yifeng Jiang}
\orcid{0009-0005-2063-8903}
\affiliation{%
 \institution{Stanford University}
 \streetaddress{450 Serra Mall}
 \city{Stanford}
 \state{CA}
 \postcode{94305}
 \country{USA}}
\email{yifengj@stanford.edu}

\author{C. Karen Liu}
\orcid{0000-0001-5926-0905}
\affiliation{%
 \institution{Stanford University}
 \streetaddress{450 Serra Mall}
 \city{Stanford}
 \state{CA}
 \postcode{94305}
 \country{USA}}
\email{karenliu@cs.stanford.edu}

\begin{abstract}
Many existing digital human models approximate the human skeletal system using rigid bodies connected by rotational joints. While the simplification is considered acceptable for legs and arms, it significantly lacks fidelity to model rich torso movements in common activities such as dancing, Yoga, and various sports. Research from biomechanics provides more detailed modeling for parts of the torso, but their models often operate in isolation and are not fast and robust enough to support computationally heavy applications and large-scale data generation for full-body digital humans. This paper proposes a new torso model that aims to achieve high fidelity both in perception and in functionality, while being computationally feasible for simulation and optimal control tasks. We build a detailed human torso model consisting of various anatomical components, including facets, ligaments, and intervertebral discs, by coupling efficient finite-element and rigid-body simulations. Given an existing motion capture sequence without dense markers placed on the torso, our new model is able to recover the underlying torso bone movements. Our method is remarkably robust that it can be used to automatically ``retrofit'' the entire Mixamo motion database of highly diverse human motions without user intervention. We also show that our model is computationally efficient for solving trajectory optimization of highly dynamic full-body movements, without relying any reference motion. Physiological validity of the model is validated against established literature.
\end{abstract}

%
%
\setcopyright{acmlicensed}
\acmJournal{TOG}
\acmYear{2023} \acmVolume{42} \acmNumber{4} \acmArticle{0} \acmMonth{8} \acmPrice{15.00}\acmDOI{10.1145/3592425}

%
%


\maketitle

\newcommand{\myparagraph}{\vspace{0.1cm}\noindent\textbf}
\def\Tabref#1{Table~\ref{#1}}

\definecolor{MyDarkBlue}{rgb}{0,0.08,1}
\definecolor{MyDarkGreen}{rgb}{0.02,0.6,0.02}
\definecolor{MyDarkRed}{rgb}{0.8,0.02,0.02}
\definecolor{MyDarkOrange}{rgb}{0.40,0.2,0.02}
\definecolor{MyPurple}{RGB}{111,0,255}
\definecolor{MyBlack}{RGB}{0,0,0}
\definecolor{MyRed}{rgb}{1.0,0.0,0.0}
\definecolor{MyGold}{rgb}{0.75,0.6,0.12}
\definecolor{MyDarkgray}{rgb}{0.66, 0.66, 0.66}
\definecolor{MyWineRed}{rgb}{0.694,0.071, 0.149}

\newcommand{\hwan}[1]{\textcolor{MyDarkGreen}{[Hwan: #1]}}
\newcommand{\yifeng}[1]{\textcolor{MyDarkOrange}{[Yifeng: #1]}}
\newcommand{\karen}[1]{\textcolor{MyDarkRed}{[Karen: #1]}}

\newcommand{\switchhwan}[1]{%
   \ifthenelse{\equal{#1}{0}}{\renewcommand{\hwan}[1]{}}{}}
\newcommand{\switchyifeng}[1]{%
   \ifthenelse{\equal{#1}{0}}{\renewcommand{\yifeng}[1]{}}{}}
\newcommand{\switchkaren}[1]{%
   \ifthenelse{\equal{#1}{0}}{\renewcommand{\karen}[1]{}}{}}

\switchhwan{1}
\switchyifeng{1}
\switchkaren{1}

\newcommand{\norm}[1]{\left\lVert#1\right\rVert}

\DeclareRobustCommand\onedot{\futurelet\@let@token\@onedot}
\def\@onedot{\ifx\@let@token.\else.\null\fi\xspace}
\def\iid{i.i.d\onedot}
\def\eg{e.g\onedot} \def\Eg{E.g\onedot}
\def\ie{i.e\onedot} \def\Ie{I.e\onedot}
\def\cf{\emph{c.f}\onedot} \def\Cf{\emph{C.f}\onedot}
\def\etc{etc\onedot} \def\vs{vs\onedot}
\def\wrt{w.r.t\onedot} \def\dof{d.o.f\onedot}
\def\etal{et al\onedot}

\newcommand{\figtodo}[1]{\framebox[0.8\columnwidth]{\rule{0pt}{1in}#1}}
\newcommand{\figref}[1]{Figure~\ref{fig:#1}}
\renewcommand{\eqref}[1]{Equation~(\ref{eq:#1})}
\newcommand{\secref}[1]{Section~\ref{sec:#1}}

\newcommand{\argmin}{\operatornamewithlimits{argmin}}
\newcommand{\sign}{\operatornamewithlimits{sign}}
\section{Introduction}
Realistic modeling and simulation of human body movements have a wide range of applications from entertainments to robotics to medicine. Conventionally, human models used in computer graphics focus on visual realism while those used in biomechanics focus on anatomical realism. With recent advances in modeling, simulation and generative models, digital humans that achieve both high fidelity in perception and in functionality, while still robust and computationally efficient, are closer to the horizon.

Human skeletal system is typically approximated as articulated rigid bodies, with hinge or ball-and-socket joints connecting the body segments. While the simplification can be acceptable for legs and arms, the human torso is significantly more complex. For example, the vertebrae bones can rotate to provide spine flexibility and can also translate to absorb compressive impact along the spinal column. Similarly, the shoulder complex manifests complicated range of motion (RoM) governed by multiple non-linear inter-bone constraints. 
For visualizing and synthesizing upper-body-rich human activities, such as gymnastics, dancing, and baseball pitching, using only rotational joints to model the human torso often leads to laborious and ad-hoc tuning in order to achieve visual plausibility. 

This paper takes a step toward building a biomechanically realistic torso model that goes beyond the chain-of-rigid-bodies structure. Instead of connected by joints, the vertebral bones of our spine model are connected and regulated by discs, ligaments and facets, where we simulate the elastic discs with finite-element method based on Projective Dynamics \cite{bouaziz2014projective}, two-way coupled with the bones, and simulate the ligaments and facets as non-linear inequality constraints in an LCP (Linear Complementary Problem) framework. Similarly, the movements of shoulder complex are also subject to their own non-linear constraints. The interplay of all elements collectively determines the RoM of the skeleton, as well as its dynamic characteristics. The model parameters are systematically adjusted such that the simulated torso can match the real human's RoM.


Our torso skeleton is kinematically and dynamically accurate and easily swappable with the simplified torso of existing human models. We show that existing motion capture (moCap) data can be "retrofit" to obtain realistic and detailed torso and shoulder bone movements that match the movements of limbs in the moCap data. Because this problem is inherently under determined, the dynamic constraints imposed by our model are necessary for solving physically and anatomically plausible torso motions. As such, our method can serve as a principled tool to enrich motion capture data by adding unobserved skeletal movements. We also show that our model is computationally fast enough for optimal control with physics simulation. We optimize a sequence of control force from the torso to produce plausible motions, without using any reference data. In comparison to simplified torso, our model can generate more momentum without anatomically implausible artifacts, achieving more human-like movements. 


We evaluate our model by reaffirming findings from biomechanics literature, demonstrating the biological validity of our model and its potentials to be used for biomedical analysis. The detailed analysis shows that the RoM of our spine model is similar to those reported in biomechanics literature. We also found that an emergent behavior of our torso model consistent with the well-known Scapulohumeral rhythm, a key signature movement of human shoulder complex.

\section{Related Work}

Existing human models for character animation often depend on arbitrary parameterization and manual rigging. Artists usually design human models to have sufficient degrees of freedom {(Dofs)} for reconstructing captured human motions \cite{kovar2008motion, holden2017phase}, but anatomically realistic range of motion and physical properties are rarely top priories. Automatic rigging methods have been proposed to accelerate the manual process by learning rig parameters from a collection of motion data \cite{le2014robust, hong2010data,xu2020rignet}. Our model can also be used as an animation rig, but we differ from existing methods in that we utilize biomechanics knowledge and physics simulation to synthesize additional details of the motion that cannot be observed by the moCap device, rather than simply fitting to the observed signals using arbitrarily parameterized rig. 

Physics-based modeling of human body parts has been broadly investigated in computer graphics, such as {fingers \cite{sachdeva2015biomechanical}}, face \cite{sifakis2005automatic}, hair \cite{selle2008mass}, muscles \cite{angles2019viper}, eyes \cite{nakada2018deep}, jaws \cite{daumas2005jaw}, skin \cite{mcadams2011efficient}, and feet \cite{camacho2002three}. For full human skeletons, however, most existing methods adopt the structure of a rigid-body tree connected by standard hinge or ball-and-socket joints, despite that rotation joints are known to be error-prone for modeling certain human parts, such as the shoulder complex \cite{maurel2000human}. The most closely related work is by Lee and Terzopoulos ~\shortcite{Lee2006HeadsUp}, who built a neck model connected with vertebrae bones using ball-and-socket joints. In their work, pivot points of each joint are carefully selected following biomechanics literature \cite{kapandji1974physiology}. { Lee and Terzopoulos~\shortcite{lee2008spline} also introduce spline joints to model complex biomechnical joints such as the scapulothoracic joints and the knee.} Rotational spring-dampers are designed to mimic the effect of intervertebrae discs. The same chain-of-rigid-bodies framework is later extended to a full upper-body model in \cite{lee2009upperbody}, combined with finite-element-based muscle and soft tissue modeling. Our model does not consider muscles. We focus on detailed and realistic modeling of skeletal connections between vertebrae bones, the rib cage, and the shoulder complex, allowing more anatomically correct range of motion. 

While our work focuses on the modeling of human, control is also an inseparable component in physics-based animation \cite{witkin1988spacetime, yin2007simbicon, peng2018deepmimic}. Multiple previous studies have revealed that better modeling can lead to more effective control. For instance, more humanlike behaviors were observed when simulated muscles \cite{komura2000creating, wang2012optimizing}, learned actuators \cite{jiang2019synthesis}, or passive dynamics of joints \cite{liu2005learning} were used. Our work also shows that a detailed spine model enables natural torso motion to emerge with little tweaking on the optimization cost function.

A rich body of work in biomechanics and medicine has been focusing on modeling human spines. {Gray and Norman~\shortcite{monheit1990kinematic} propose a kinematic model of the human spine with all vertebral joint movements having three axes of rotations.} Panjabi and colleagues \shortcite{PANJABI1976stiffness} modeled the discs as bushing elements between vertebrae bones, parameterized as $6$ by $6$ stiffness matrices. Such bushing elements have been extended to have non-linear stiffness to better reflect the complex dynamics of intervertebrae discs, with parameters identified from in-vitro experiments \cite{wang2020implementation}. 
Dicko and colleagues \shortcite{Dicko2015Construction} modelled discs using finite-element and ligaments as non-linear springs. However, their model was only used for static spine pose analysis, possibly due to less stable dynamics caused by springs. In general, robustness and computational efficiency are the main reasons why most existing spine models were rarely used beyond static pose equilibrium analysis \cite{wang2020implementation, GHEZELBASH2018Effects}. Our model adopts a hybrid FEM plus rigid approach, utilizing an efficient finite-element simulation framework, coupled with efficient rigid-body simulation with robust impulse-based constraint handling through LCP solvers, enabling stable simulation of full-body motion even with large foot-ground impact.

Human shoulder complex has also been studied extensively in biomechanics, Seth and colleagues \shortcite{seth2016biomechanical} showed that analytic modeling of the scapulothoracic joint improved the accuracy of shoulder movement reconstruction. They used measured statistics of the scapulothoracic joint movements gathered from in-vivo experiments~\cite{mcclure2001direct}. More realistic skeleton modeling facilitated the construction of shoulder muscle models in their later study \cite{seth2019muscle}. Biomechanics researchers have also combined spine and shoulder models in biomechanics simulators, e.g. OpenSim \cite{delp2007opensim}, to study various activities. For example, \cite{Stenteler2016force, Bruno2015Development} studied spine load and reaction forces during primitive motions such as bending and lifting with weight, and \cite{cazzola2017cervical} studied shoulder injury of rugby players.

\section{Torso Modeling}

\begin{figure}
    \centering
    \includegraphics[width=0.95\linewidth]{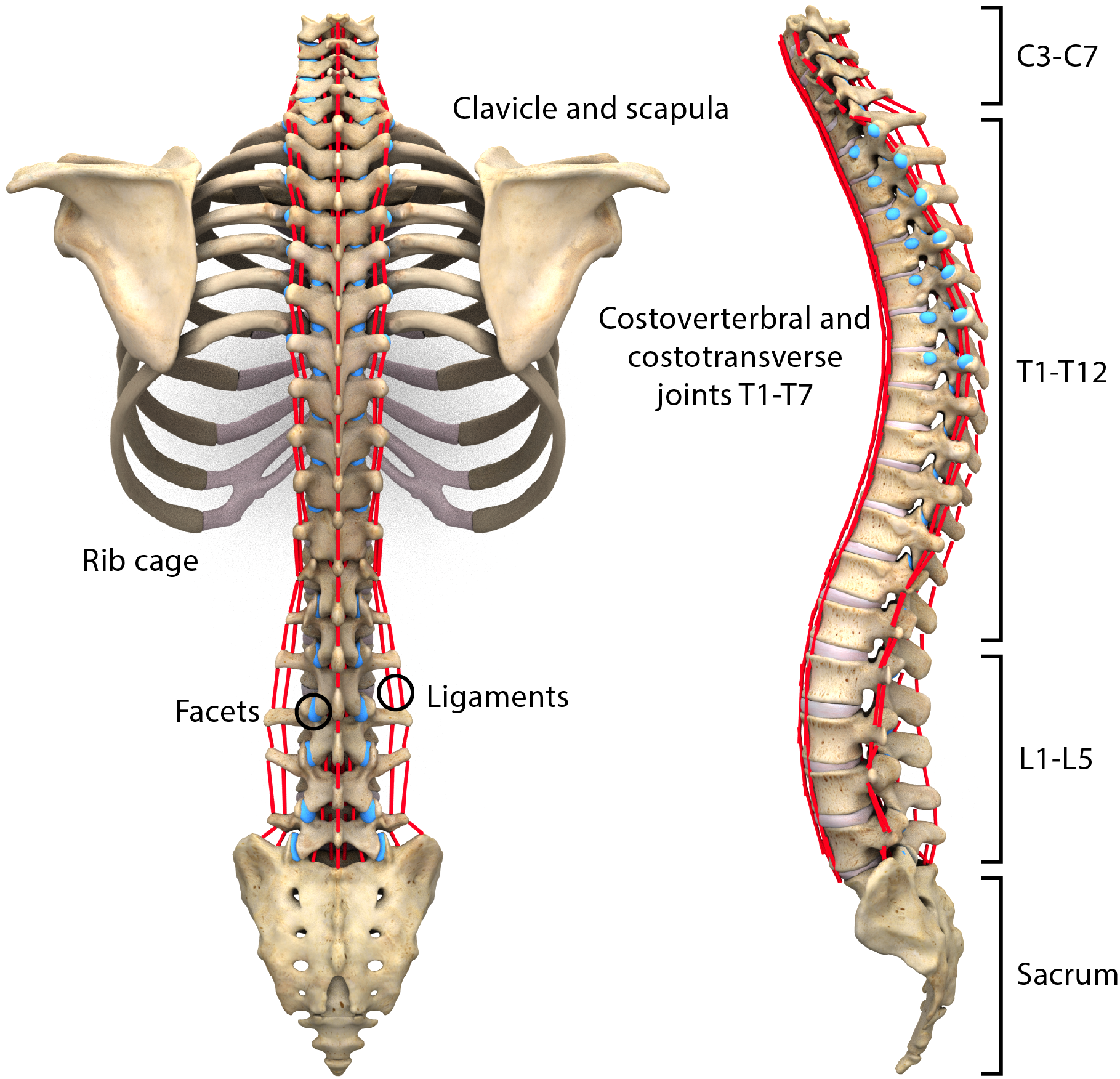}
    \caption{\label{fig:character} Our torso model include verterbrae bones, discs, ligaments, facet joints, a rib cage, and a clavicle and a scapula on each side of the shoulder complex.}
\end{figure}
We present a physics-based model for human torso that simulates detailed human anatomy. Our torso model consists of bones, joints, discs, ligaments, and facets. We include 23 vertebrae bones (from sacrum to C3) assembling a spinal column, a rib cage, and a clavicle and a scapula on each side of shoulder complex (See Figure~\ref{fig:character}). 20 ligaments and 44 facets are attached to the spine, and 23 vertebrae discs are sandwiched between adjacent vertebrae bones. There are 28 additional costovertebral and costotransverse facets joints that connect the rib cage to the thoracic bones. The collaboration of the ligaments, the discs and the joints defines the passive dynamics of our model as well as kinematic characteristics such as the range of motion. Muscles are not included in our current model.

All the bones are modeled as rigid bodies with detailed geometries used by anatomical modeling~\cite{zygote1994}. The rib cage is assumed to be a single rigid body as the movements of the ribs are mostly associated with breathing. We use ball-and-socket joints to connect the rib cage and the clavicle, and the clavicle and the scapula. Between the vertebrae bones and between the rib cage and the thoracic vertebrae, we connect them with placeholder 6-DoF free joints, and the relative movements are instead constrained by finite-element simulation of the discs and non-linear constraints rising from ligaments and facets. The use of 6-DoF joints allows us to still treat the torso as an articulated rigid-body tree with $\bm{q}$ being the state of the rigid-body in generalized coordinates, facilitating incorporating existing frameworks for physics simulation. Two-way weak coupling between the finite-element discs and the rigid-body skeleton is discussed in Section \ref{sec:disc} and \ref{sec:dynamics}. 


\begin{figure}
    \centering
    \includegraphics[width=0.9\linewidth]{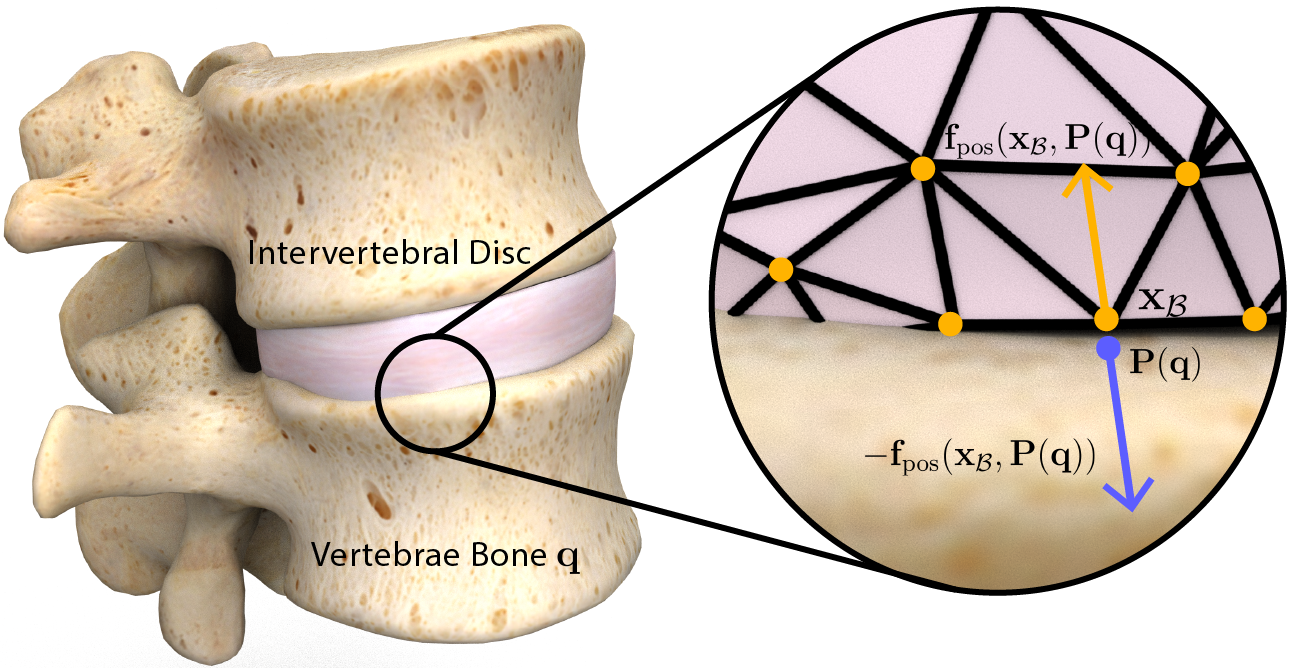}
    \caption{\label{fig:vertebrae_bone} We simulate intervertebral discs as deformable materials using finite element method. }
\end{figure}

\subsection{Intervertebral Disc} 
\label{sec:disc}
Intervertebral discs are deformable materials attached in-between adjacent vertebrae bones (See Figure~\ref{fig:vertebrae_bone}). The elasticity of the discs prevents the bones from colliding and rubbing against each other, and absorbs and dissipates large compressive impacts on the spinal column. Such a deformable body is often simulated using Finite Element Method (FEM) which discretizes a body into a finite number of elements such as tetrahedrons. The states of the disc at time $t$ can be represented by its nodal positions $\bm{x}_t \in \mathbb{R}^{3k}$, where $k$ is the number of discretized vertices.

Following previous work, we adopt the standard implicit Euler integration in its variational form \cite{gast2015optimization, martin2011example}, where at each time step $t$, we solve an non-linear optimization problem for the current state $\bm{x}_t$:

\begin{equation}
\label{eq:integrator}  
\bm{x}_t = \argmin \frac{1}{2h^2} \norm{\bm{M}^{\frac{1}{2}} (\bm{x}_t - \bm{x}_{t-1} - h\bm{v}_{t-1})}_F^2 + \sum_i k_iE_i(\bm{x}_t).
\end{equation}

Here $\bm{M}$ is the mass matrix and $h$ is timestep size. $\bm{v}_t$ is obtained through backward finite differencing of $\bm{x}_t$. By minimizing Equation \ref{eq:integrator}, we essentially solve for $\bm{x}_t$ that balances the change of momentum and the forces expressed as the gradients of the potential energies $E_{i}$, with scalar weights $k_i$. For our case, $E_{i}$ include isotropic elasticity \cite{chao2010simple} and volume preservation \cite{bouaziz2014projective} of the disc.


The minimization problem can be solved by modified Newton's Method, but it is usually computationally expensive. This is because Hessian $\nabla^2E_i(\bm{x})$ obtained through linearization changes at each time step, preventing us from utilizing matrix factorization to accelerate. Facing this issue, Projective Dynamics \cite{bouaziz2014projective} showed that certain types of energy functions can be solved much more efficiently in an alternating "local-global" manner. Following their  framework, we define each $E_{i}$ as a minimization process:
\begin{equation}
\label{eq:pd-energy}  
E_i(\bm{x}) = \min_{\bm{m}_i \in \mathcal{C}_i}\frac{1}{2}||\bm{A}_i\bm{x} - \bm{m}_i||^2,
\end{equation}
where the constraint manifold $\mathcal{C}_i$ and coefficient matrix $\bm{A}_i$ depend on the specific energy term. Equation \ref{eq:pd-energy} solves for $\bm{m}_i$, the projection of $\bm{A}_i\bm{x}$ onto $\mathcal{C}_i$. The projection error is then defined as the energy $E_i$. Since each $E_i$ is independently computed without considering other $E_i$'s, we call this step a ``local'' solve. 
A large class of FEM energies can be written into this projection form. For instance, we can cast the isotropic elasticity term as the distance from the current deformation gradient to its closest projection on the ${\rm SO(3)}$ manifold (isometry rotation with zero deformation), such that a rigid rotation of the rest-pose tetrahedron will lead to zero elastic energy.

Once all $\bm{m}_i$'s are obtained through local solves, we proceed with the standard Newton steps (``global solve''). We plug in the solved $\bm{m}_i$ in Equation \ref{eq:pd-energy} and express $E_i$ as a quadratic function in $\bm{x}$, where the constant $\bm{A}_i$ enables acceleration using Cholesky decomposition. In practice, since the discs are tightly attached to the bones with negligible inertial effects comparing to the bones, we adopt a quasi-static simplification, effectively forgoing the momentum term ($h \rightarrow \infty$) to improve simulation stability, similar to previous works \cite{lee2018dexterous}. { As such,  Equation~\ref{eq:integrator} simplifies to: $\bm{L}\bm{x}_t = \bm{d}$, where $\bm{L} = \sum_i k_i \bm{A}_i^{\rm T}\bm{A}_i$ and $\bm{d} = \sum_i\bm{A}_i^{\rm T}\bm{m}_i$.}

An additional energy term $E_{\rm pos}$ is needed to couple the vertebrae bones and the discs. We compute a set of attached points between the bone surface and the disc surface at the rest pose and enforce the attachment via minimizing $E_{\rm pos} = \norm{\bm{x}_\mathcal{B} - \bm{P}(\bm{q})}^2$, where $\bm{P}(\bm{q})$ evaluates attach points on the bone surface in the world frame given the current state $\bm{q}$, and  $\bm{x}_\mathcal{B}$ are the boundary disc vertices in the world frame. Note that we use the bone configuration $\bm{q}$ from the previous time step, resulting in a weak coupling between the bones and the discs. $\bm{f}_{\rm pos} := -\nabla E_{\rm pos}$  is therefore the coupling force between finite-elements and rigid bodies (See Figure \ref{fig:vertebrae_bone}).




\begin{figure}
    \centering
    \includegraphics[width=0.95\linewidth]{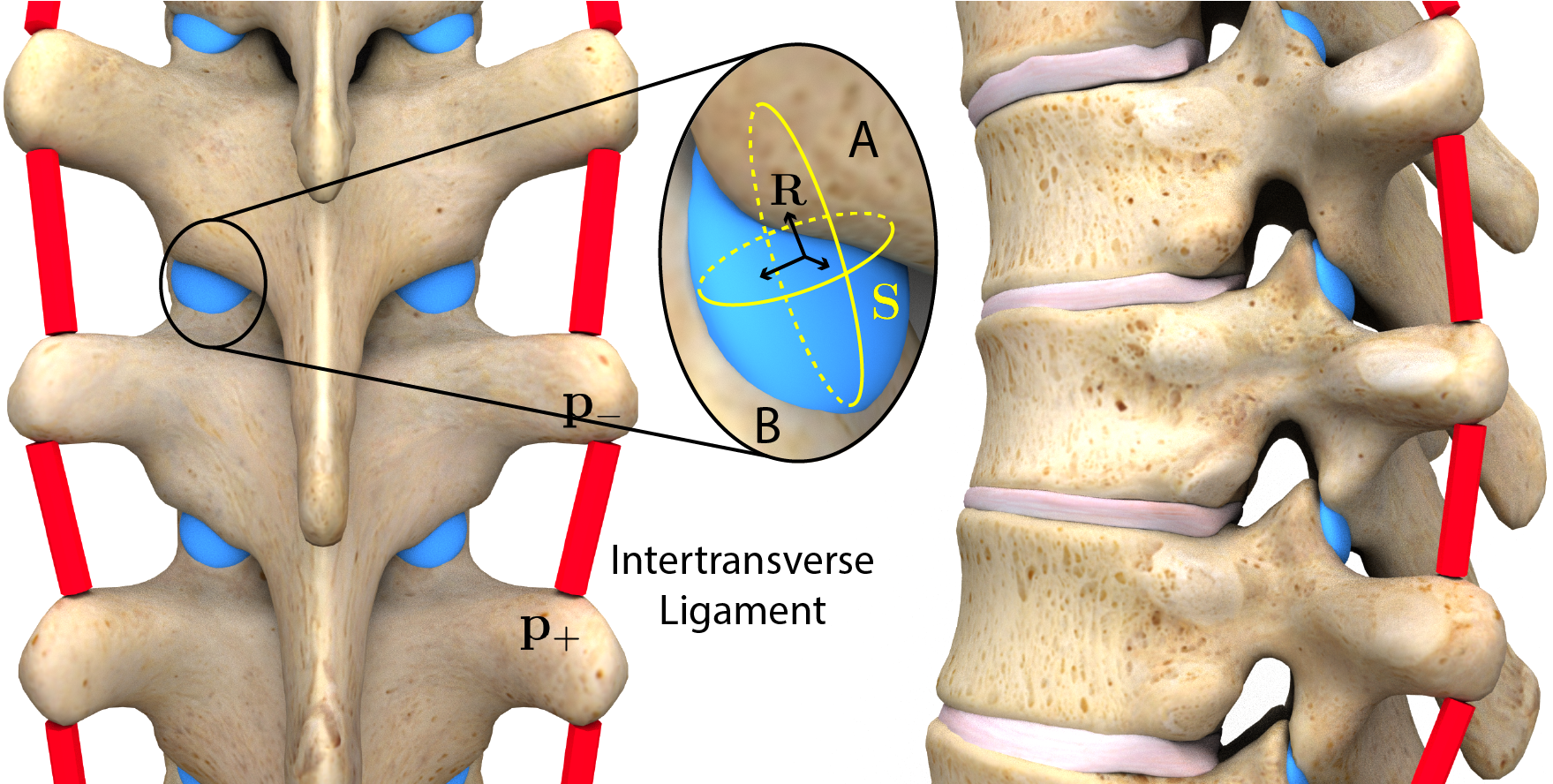}
    \caption{\label{fig:ligament_and_facet} Ligaments (red) and facet joints (blue). $\bm{p}^+$ and $\bm{p}^-$ being waypoints on the ligament's routing from its origin on sacrum to insertion on neck. The location on cartilage A where the facet resides is termed "inferior articular process" and the counterpart on cartilage B is termed "superior articular process".}
\end{figure}

\subsection{Ligament} A ligament is a spring-like fibrous connective tissue attached between bones. It regulates the relative movements among the bones, keeping them stable. The anatomical functionality of each ligament varies by attachment sites (termed \emph{origin} and \emph{insertion}). For example, the intertransverse ligament is routed on the transverse process of the vertebrae bone, and it mainly limits the opposite-side lateral flexion of the spine (See Figure~\ref{fig:ligament_and_facet}).
We use piece-wise linear approximations to represent the ligament and its geometry. Starting from the origin of a ligament, we divide the ligament into multiple line segments, where each segment is expressed by $\bm{p}^+$ and $\bm{p}^-$ defined in the local coordinate of the associated bones respectively.

Each ligament is represented by a unilateral spring; it generates contractile forces when the ligament is extended and zero force when relaxed, a characteristic described by the Hill's muscle model \cite{zajac1989muscle}. {According to the model, the force-length curve of ligaments has two phases: tendon elongations below the thresholds create negligible forces, while tendon elongations above a threshold create exponentially growing forces.} In practice, it is desirable to avoid a large amount of contractile forces as it decreases simulation stability. Hence we add inequality constraints (to be solved by a stable LCP time-stepping scheme) in our simulation where they preemptively bound the length of the ligament within a moderate range:
\begin{equation}
C_{\rm ligament}(\bm{q}) = \|\bm{p}^+(\bm{q}) - \bm{p}^-(\bm{q})\|^2 \leq l_{\rm max},
\end{equation}
where $l_{\rm max}$ is the maximum length that varies by each ligament (see Section \ref{sec:param-id}). {Comparing with the Hill's model, this modification to inequality constraints (zero force below threshold and infinite-as-needed beyond threshold) doesn’t change the biomechanical property much.} Note that one ligament will create multiple $C_{\rm ligament}(\bm{q})$ constraints with different $l_{\rm max}$ for each segment. Section \ref{sec:param-id} will describe how the ligament parameters are identified systematically.

\subsection{Facet}  The facets are located on the posterior side of the vertebrae bones, as well as between the rib cage and thoracic vertebrae bones. They are synovial joints consisting of two cartilages facing each other, wrapped up by the joint capsules. One cartilage (A) slides on the surface of the other cartilage (B) with near-frictionless movement due to highly smooth surface and the existence of fluids (See Figure~\ref{fig:ligament_and_facet}). Facet forms a closed-chain kinematics with the disc which increases the stability and robustness of the spinal column against excessive external perturbations. The surface geometry of the cartilage B determines the movements of the facet~\cite{williams2010shape,drake2009gray}. As the cartilage is flat and oval-shaped, we represent the surface geometry of cartilage B as a flat ellipsoid:
\begin{equation}
\label{eq:elipsoid}
C_\text{facet}(\bm{p}) = ||\bm{S}^{-1}\bm{R}^{\rm T}(\bm{p} - \bm{p}_{\rm center})||^2 - 1,
\end{equation}
where $\bm{S}\in\mathbb{R}^{3 \times 3}$ is the diagonal matrix representing each axis scale, $\bm{R}\in{\rm SO(3)}$ is the rotation matrix where each column denotes the axis of the ellipsoid, and $\bm{p}_{\rm center}$ is the center of the ellipsoid. Here, the level-set $C_{\rm facet}(\bm{p}) = 0$ 
represents the surface where we would like points on cartilage A to stay on. In other words, we need to select three non-colinear points on A that are also on $C_{\rm facet}(\bm{p}) = 0$ at the rest configuration, and the constraint solver will then keep these three "wheels" on B's ellipsoid surface during the entire simulation. In practice, we select four points instead of three to ensure numerical robustness, and relax $C_{\rm facet}(\bm{p}) = 0$ to $|C_{\rm facet}(\bm{p})| \leq 0.2$.

\begin{figure}
    \centering
    \includegraphics[width=0.9\linewidth]{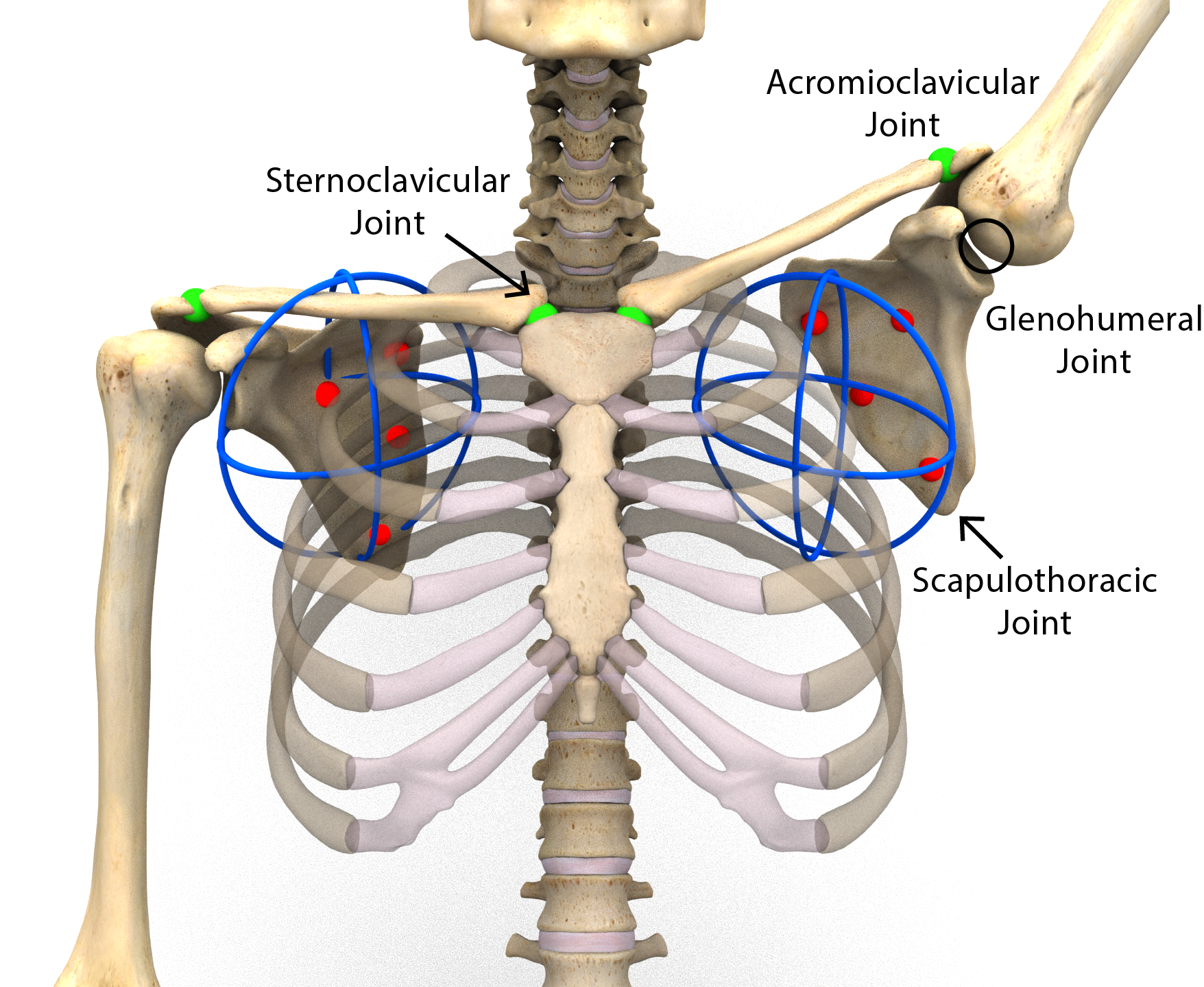}
    \caption{\label{fig:shoulder} The anatomy of shoulder joint. We use ball-and-socket joints for the sternoclavicular joint, acromioclavicular joint, and glenohumeral joint (green). We utilize facet constraints to simulate the scapulothorcic joint where the ellipsoid (blue) defines the surface of the rib cage whereas points attached to the scapula (red) are moving on the ellipsoid.}
\end{figure}
\subsection{Shoulder}

We also utilize the ellipsoid constraint used for facets to simulate the shoulder movements (See Figure~\ref{fig:shoulder}). The clavicles and the scapula together determine the location and the orientation of the shoulder. The joint of the clavicle is on the sternum (sternoclavicle joint) and can be modeled by a rotational joint. The movement of the scapula is much more complex and determined by both the acromio-clavicle joint and the scapulothoracic joint. The scapulothoracic joint represents the sliding of the scapula on the posterior of the rib cage. It has been revealed in many studies (e.g. \cite{seth2016biomechanical}) that the rib cage can be approximated by an ellipsoid on which the scapula slide on. This movement is identical to the movement of the facet joint so we add constraints between the rib cage and the scapula to mimic the scapulothoracic joint anatomy. Unlike facets joint, however, the ellipsoid constraints for the scapula will create a close-chain kinematics.

\subsection{Torso Dynamics}
\label{sec:dynamics}
With finite-element simulation presented in Section \ref{sec:disc}, the equations of motions of our rigid skeleton model can be described by the Lagrangian dynamics, where we solve for $\ddot{\bm{q}}$ to drive the simulation forward:
\begin{flalign}
\label{eq:eom}
&\bm{M}(\bm{q})\ddot{\bm{q}} + \bm{c}(\bm{q}, \dot{\bm{q}}) = \bm{\tau}_{\rm disc}(\bm{x}_\mathcal{B},\bm{P}(\bm{q})) + \bm{\tau}_{\rm ext} \;\;\; {\rm subject\;to} \\
\label{eq:c_ligament}&\bm{C}_{\rm ligament}(\bm{q}) \leq \bm{l}_{\rm max}\\
\label{eq:c_facet}&|\bm{C}_{\rm facet}(\bm{p}(\bm{q}))| \leq 0.2,
\end{flalign}
where $\bm{q}$ is the generalized positions, $\bm{M}(\bm{q})$ is the mass matrix, $\bm{c}(\bm{q}, \dot{\bm{q}})$ is the coriolis and gravitational forces, $\bm{\tau}_{\rm ext}$ is the external forces, and $\bm{\tau}_{\rm disc}(\bm{x}_\mathcal{B}, \bm{P}(\bm{q})) = -\bm{J}^{\rm T}\bm{f}_{\rm pos}$ is the sum of forces generated by the disc attached to the bones via positional constraints $E_{\rm pos}(\bm{x}_\mathcal{B}, \bm{P}(\bm{q}))$. $\bm{J}$ is the Jacobian matrix that maps the generalized coordinate to the attachment (bone surface) points. Equation \ref{eq:c_ligament} and \ref{eq:c_facet} are the ligament constraints and facet joint constraints, respectively. 



Algorithm~\ref{alg:forward_dynamics} shows the process of forward dynamics of our model for each time step $t$, where we sequentially solve the FEM simulation and the articulated rigid body simulation. For FEM, we iterate between solving the local projection for each constraint (Equation \ref{eq:pd-energy}) and linear solves for the Newton steps. Featherstone's algorithm \cite{featherstone2014rigid} is used to solve Equation \ref{eq:eom} while the constraints~\ref{eq:c_ligament} and \ref{eq:c_facet} are solved via impulse-space simulation which yields a Linear Complementary Proplem (LCP)~\cite{anitescu1997formulating}. We use Danzig's LCP solver \cite{cottle2009linear} to compute the lagrangian multipliers of the constraints which are eventually used to modify the generalized velocities $\dot{\bm{q}}_{t+1}$ such that they will not violate the constraints. Finally, the solved $\bm{q}_{t+1}$ will be used for the FEM simulation at next time step.


\begin{algorithm}
\caption{Torso Forward Dynamics}\label{alg:forward_dynamics}
{
    Input: ${\rm Disc\;Positions}$ $\bm{x}_{t}$, ${\rm Joint\; Positions, Joint \;Velocities}$ $\bm{q}_{t},\dot{\bm{q}}_{t}$, ${\rm Bone\;Surface\;Locations}$ $\bm{P}(\bm{q}_t)$ $\bm{L} = \sum_i k_i\bm{A}_i^{\rm T}\bm{A}_i$\\
    \For{$l=0,1,\cdots,{\rm numFEMIterations}$}
    {
        
        $\bm{d} = \bm{0}$ \\
        \For{$i=0,1,\cdots,{\rm numFEMEnergies}$}
        {
            $\bm{m}_i = {\rm project}(\bm{A}_i\bm{x}_t ,\mathcal{C}_i)$\\
            $\bm{d} = \bm{d} + \bm{A}_i^{\rm T}\bm{m}_i$\\
        }
        $\bm{x}_t = \bm{L}^{-1}\bm{d}$ \\
    }
    $\tau_{\rm disc} = -\bm{J}^{\rm T}\bm{f}_{\rm pos}({\bm{x}_t}_{\mathcal{B}}, \bm{P}(\bm{q}_t))$\\
    $\ddot{\bm{q}}_{t+1} = {\rm ForwardDynamics}(\tau_{\rm ext}, \tau_{\rm disc}, \bm{q}_t, \dot{\bm{q}}_t)$\\
    $\dot{\bm{q}}_{t+1} = {\rm integrateVelocities}(\ddot{\bm{q}}_{t+1}, \dot{\bm{q}}_{t})$\\
    $\dot{\bm{q}}_{t+1} = {\rm ResolveConstraintsByLCP}(\dot{\bm{q}}_{t+1}, \bm{C}_{\rm ligament}(\bm{q}_t),\bm{C}_{\rm facet}(\bm{q}_t))$\\
    $\bm{q}_{t+1} = {\rm integratePositions}(\dot{\bm{q}}_{t+1}, {\bm{q}_{t})}$\\
    $\bm{x}_{t+1} = \bm{x}_t$\\
}

\end{algorithm}

\begin{figure}
    \centering
    \includegraphics[width=0.8\linewidth]{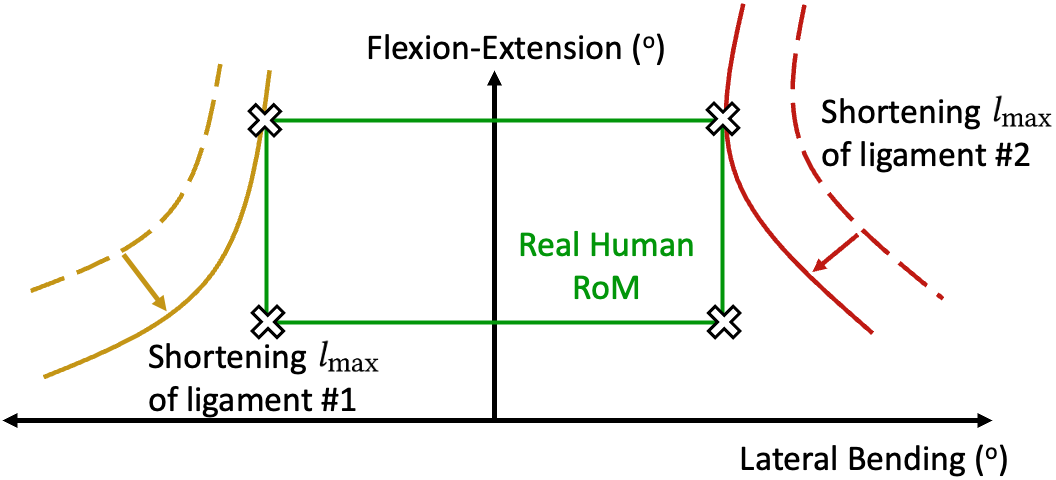}
    \caption{\label{fig:rom_optimizer} The maximum ligament lengths are adjusted sequentially to match the human's range of motion. The cross symbols represented the extreme relative orientations between two verterbrae bones known to biomechanics literature. Ligament 1 is shortened until one of extreme poses is no longer feasible, followed by the repeated operation for Ligament 2. Axial rotation DoF not shown here for simplicity.}
\end{figure}
\subsection{Parameters Identification}

\label{sec:param-id}
\paragraph{Ligaments} The maximum lengths $l_{\rm max}$ of the the ligaments are determined from biomechanics data of spine range of motion. For each adjacent pairs of vertebrae bones, existing literature \cite{savlovskis-no-date} measured a few extreme relative orientations (shown as cross symbols in Figure \ref{fig:rom_optimizer}) that an average subject is able to achieve. As shown in Figure \ref{fig:rom_optimizer}, we gradually shorten $l_{\rm max}$ until it is no longer enough to realize all documented extreme poses. At each shortening step, this check is done by running the simulation for the bone pair with random actions. After all ligaments are "tightened", we will obtain a non-linear range of motion (the yellow and red lines) that is strictly more flexible than the simplex (green lines) of measured extreme poses. In practice, some of the ligaments only affects one degree of freedom, and we shorten these independent ligaments first. Though different orders of shortening can lead to slightly different non-linear shapes of the RoM, they are equally valid in the sense of matching all available measurement data.

\paragraph{Facets} The location $\bm{p}_{\rm center}$, orientation $\bm{R}$ and dimension $\bm{S}$ of each facet ellipsoid are determined by annotating the bone geometries. For each pair of the adjacent bones, we visually inspect and manually annotate the positions of \emph{superior articular process} and \emph{inferior articular process} with knowledge of anatomy. We determine the parameters such that the ellipsoid contains all the vertices on the \emph{superior articular process}. We randomly select four non-colinear points $\bm{p}$ among the vertices on the \emph{inferior articular process}. { Note that these four points are irrelevant to the shape of the facet ellipsoid and thus do not affect RoM; they are only used to enable adjacent bones to slide with each other. Therefore, randomly choosing four non-colinear points is sufficient to guarantee sliding.}

\paragraph{Discs} FEM parameters such as the Young's modulus are tuned manually for stable and visually plausible results. We leave matching disc properties to the non-linear stiffness observed in measurement data \cite{wang2020implementation} to future work.


\section{Experimental Results}
We evaluate our model based on three aspects. \emph{Realism:} Can our model generate detailed torso motions that are visually realistic and consistent with the realistic human range of motion? \emph{Robustness:} Can we use our model as an automatic tool to robustly process large-scale motion datasets? \emph{Efficiency:} Is our model computationally efficient enough for physics simulation and optimal control applications? We describe a few experiments to validate these aspects below and provide a supplemental video for further evaluation.

\subsection{Experiment setup}

Our model includes 24 place-holder free joints and 17 ball-and-socket joints, totaling $195$ DoFs. We model an adult human with a height of 170cm and a mass of 72kg. {{The mass of a bone is computed by the bone density and the bounding box of the mesh while the mass of some bones with concave shapes such as ribs, scapula, and clavicle, is manually assigned. We apply joint limits to the sternoclavicular joint, the acromioclavicular joint, and the glenohumeral joint~\cite{magee2013orthopedic} while the other joints have no limits.}
We also use passive spring-damper for all joints which increase simulation stability as well as the robustness of retrofitting. { We use a PD controller for each DoF of our torso model.}

Our model consists of 20 ligaments and 74 facets, including 44 facet joints, 28 costovertebral and costotransverse joints, and 2 scapulothoracic joints. Each ligament is modelled by a sequence of line segments, each of which corresponds to each individual constraint equation~(\ref{eq:c_ligament}). This yields 440 tendon constraints on the spinal column. We also sample four non-colinear points for each facet. As a result, our rigid-body world is simulated by 680 constraints. Our rigid-body simulation is written in C++ building upon the open-source library DART~\cite{Dart}. { We directly use the LCP solver in the DART engine for body-ground contact.} The simulation runs at 600Hz for retrofitting and 1800Hz for trajectory optimizations. Simulation runs 8 to 40 times slower than real-time depending on configurations. We tetrahedralize the disc meshes using TetWild~\cite{hu2018tetrahedral}. Each disc consists of 127 to 285 nodal vertices and 307 to 1090 tetrahedrons depending on its shape. 



\subsection{Motion Retrofitting}


Our method provides a tool to estimate detailed movements of torso bones unobserved by standard moCap devices. We apply our tool to a large existing motion capture dataset, Mixamo, without user intervention to demonstrate its robustness. We will host a large-scale public dataset of motion capture data with enriched torso movements.


Our tool takes as input a target motion on any rigged skeleton and first retarget it to a temporary skeleton, followed by simulating the detailed torso movements on our detailed skeleton. In the retargeting step, we utilize the IK-based retargeting algorithm provided by {\em MotionBuilder$^\mathrm{TM}$} to match the limb and head movements of the target motion on our temporary skeleton. The limbs and head of the temporary skeleton are identical to those of the detailed skeleton, but the torso is simplified to three ball joints, similar to many existing human rigs used in computer animation. { The temporary skeleton has shoulders that include sternoclavicular and acromioclavicular joints, as it needs to allow movements such as shoulder shrugging and straightening.} We set a wider range of motion for the temporary torso so that the arms and the head can match the target motion precisely. 


For each frame, using the current positions of the temporary skeleton, we set the positions of humerus, femur, and head bones as positional constraints and run our torso dynamics (Section \ref{sec:dynamics}) until convergence to a static equilibrium. The dynamic simulation is necessary because there are far more degrees of freedom on the torso then the constraints. In our experiments, we simulate the torso dynamics for $0.5s$ for most examples, while for some motions containing excessive bending we simulate the dynamics for $1s$.


We retrofit 1293 motion clips from {\em Mixamo} dataset including assorted dance moves, martial arts, baseball, jumping, walking, running etc. The dataset contains motion clips ranging from $1$ to $44$ seconds long, totaling an hour and 18 minutes of high quality motion data. Motion retrofitting takes 20 hours with AMD Ryzen 9 5950X. We accelerate the process with 16 parallel cores. A few representative motions can be found in the supplemental video.

\subsection{Trajectory Optimization}

We evaluate our torso model by solving trajectory optimization problems to simulate highly dynamic motions, i.e. jumping and swinging. The optimizer solves for a control trajectory to optimize the performance of a physically simulated motion sequence. Since the main goal of this experiment is to validate that the large number of DoFs does not prohibit physics simulation or optimal control in terms of efficiency, and motion naturalness is secondary, we use a very simple cost function, such as jumping as high as possible, with little regularization or tuning and without any use of human motion data.

We adopt a standard stochastic trajectory optimization framework MPPI \cite{williams2017model}, a shooting-based gradient-free algorithm for solving open-loop control sequences. Given a constant initial state $\bm{q}_0, \dot{\bm{q}}_0$, and a transition function defined by a physics simulator $\bm{q}_{t} = F(\bm{q}_{t-1}, \bm{a}_{t-1})$ with any action space $\bm{a}$, MPPI solves for ($\bm{a}_0$, $\cdots$, $\bm{a}_{N-1}$) such that the total optimal control cost $\sum_{t=0}^N c(\bm{q}_t, \dot{\bm{q}}_t)$ of the simulated trajectory $(\bm{q}_0, \bm{q}_1, \cdots, \bm{q}_N)$ is minimized. 
We additionally compare motions solved by our model to those from a fixed-torso model. Optimization takes around 30 MPPI iterations for both our model and the fixed-torso model, though our model has a much larger action space. For our model MPPI optimizations take around 6 hours each with 6 CPU cores, and the rigid-torso model is around 60x faster.

\subsubsection{Falling Test} Before running optimization, we make sure that when combining our model with an existing full-body skeleton, the combined model will be robust enough to interact with the environment and execute random control sequences commanded by MPPI during optimization. For this we drop our character from 2 meter high to the floor with small random controls commanded. As shown in the supplemental video, our model can handle discontinuous, large contact impact without issues. Interestingly, even without any optimization, our model already generates a more humanlike falling motion upon touching the ground, comparing with the fixed-torso model.

\subsubsection{Swing} In this task, the character attempts to swing on a horizontal bar from zero velocity. The objective is to maximize the product of horizontal and vertical speeds at the final frame of the trajectory, such that the character would travel the longest distance after releasing the bar:
\begin{equation}
     - \dot{q}_{y, N} \cdot \dot{q}_{z, N} + \sum_{t=0}^N 0.1 q_{x,t}^2 ,
\end{equation}
where ${q}_{x, t}, {q}_{y, t}, {q}_{z, t}$ are respectively the $x, y, z$ components of the center of mass of the character at time step $t$. The second term penalizes deviation from the sagittal plane: 

The action space at each control step is the change of PD targets of relative bone orientations. While this action space is largely redundant compared with the real human musculoskeletal system which uses about $30$ muscles to actuate the spine, we did not find MPPI having practical difficulty converging to a solution, nor did we find unnatural spine poses exhibted in the solution (except the task-irrelevant neck bones tends to be moderately jittery). We hypothesize that the coupling of discs, ligaments and facets in our spine model effectively restrict the search space of MPPI to a biomechanically realistic state-action space. On the other hand, the fixed spine character which uses only torso DoFs to swing demonstrates unsmooth and suboptimal behaviors, potentially due to its inability to utilize the spine to store and release potential energy.

\subsubsection{Jumping} The cost function tasks the character to jump as high as possible, with a small regularization term discouraging deviation from the vertical ($y$) direction:
\begin{equation}
     - \sign (\dot{q}_{y, N}) \frac{\dot{q}_{y, N}^2}{2g} - q_{y, N} + \sum_{t=0}^N 0.001 (q_{x,t}^2 + q_{z,t}^2).
\end{equation}
For the jumping task, the action space for both our model and the fixed-torso model additionally include the knee and ankle DoFs. The difference between motions solved with two models are more subtle for this task, as excepted since spine mainly performs a balancing and buffering role in jumping. Nevertheless, the motion simulated by our torso model exhibits a propagation of acceleration due to the ground reaction force from the lower limbs through the spine, while the motion simulated by the fixed-torso model only uses ankles to exert large torque, resulting in more robot-like motion.

\begin{figure}
    \centering
    \includegraphics[width=0.9\linewidth]{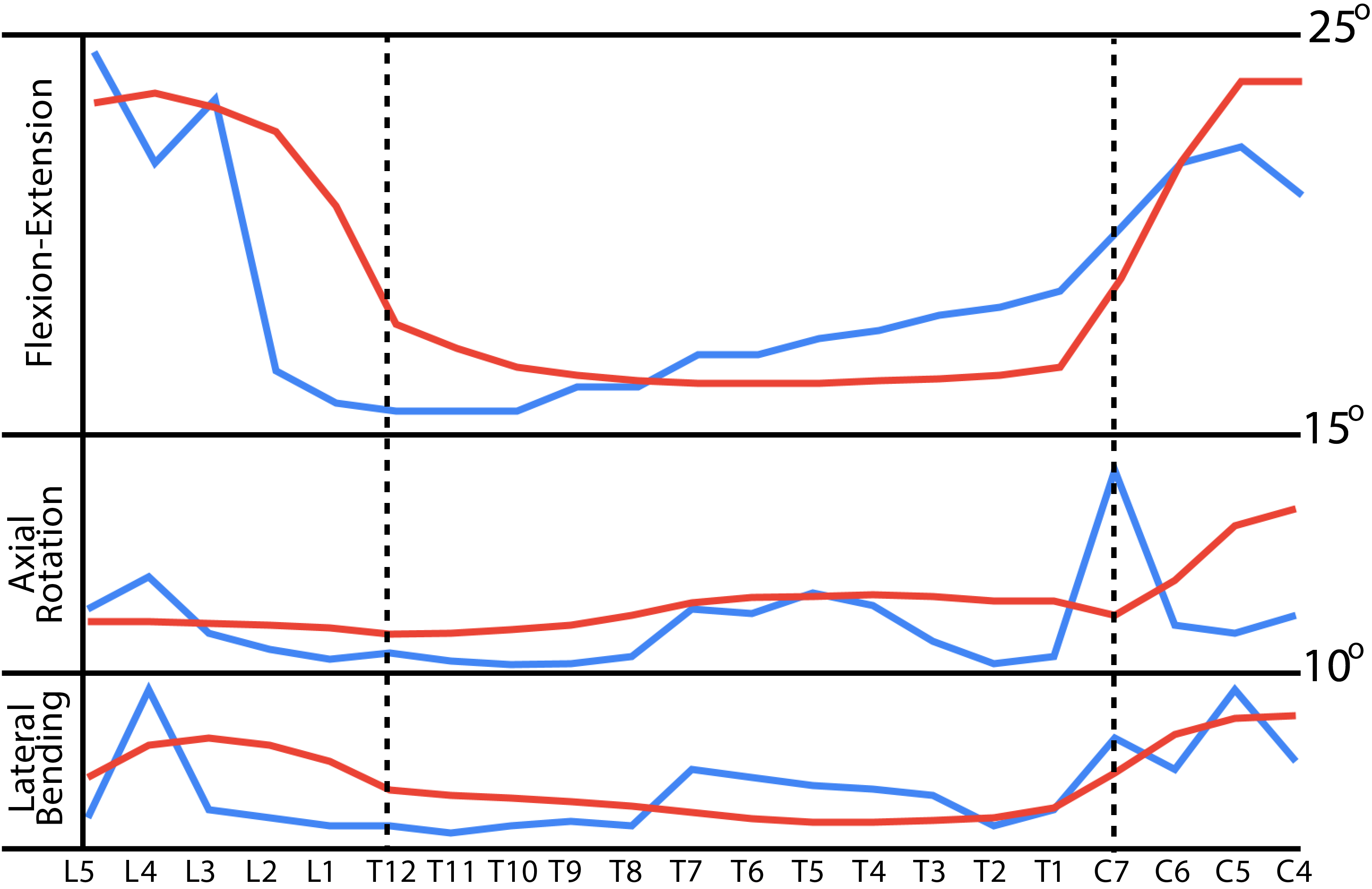}
    \caption{ The ranges of motion of the vertebrae bones. (red) Real human RoM of each bone. (blue) RoM measured by our model.}
    \label{fig:rom}
\end{figure}
\subsection{Comparisons with Real Human Data}
\subsubsection{Range of Motion}

An effective validation of our model is to compare the range of motion (RoM) of each vertebrae bone in the simulated motion against real human RoM recorded in biomechanics literature. Because we do not explicitly ``set'' joint limits for our torso model, the resulting RoM is an emergent quantity from the complex interplay of physical and geometry properties of discs, ligaments, and facet joints.

We compare the range of motion of each bone with established biomechanics literature~\cite{anderst2015three, bhalla1969normal,fujii2007kinematics}. The RoM of existing real human data is highly variant by subject's age, sex, and physical capability. We compare the RoM of our model with the average of existing data shown in red line in Figure~\ref{fig:rom}). For each vertebrae bone, we measure the maximum and the minimum angle along anatomical axes (Flexion-Extension, Axial rotation and Lateral bending) to identify the RoM. The ranges of axial rotation and lateral bending are divided by half as they are symmetric. The RoM of our model is measured by the minimum and maximum angles (5\% outliers are removed) among all retrofitted sequences of the Mixamo dataset.

We observe that most verterbrae bones have similar RoMs compared to the biomechanics literature. There are still a few noticeable differences. For flexion-extension, the real human has a wider RoM on L1 and L2 while our model has a wider RoM on upper thoracic region. We suspect that the maximum length of ligaments we identified might be larger than the real human statistics due to our overly conservative algorithm. As a result, bending upper thoracic region is sufficient to match the positional constraints. We also observe a peak on C7 especially in axial rotation. We suspect that the locations and the shapes of the costovertebral joints and the costotransverse joints overly constrain the thoracic bones, and thus C7 has to be more flexible to compensate the narrowed RoM on the thoracic bones.
\begin{figure}
    \centering
    \includegraphics[width=\linewidth]{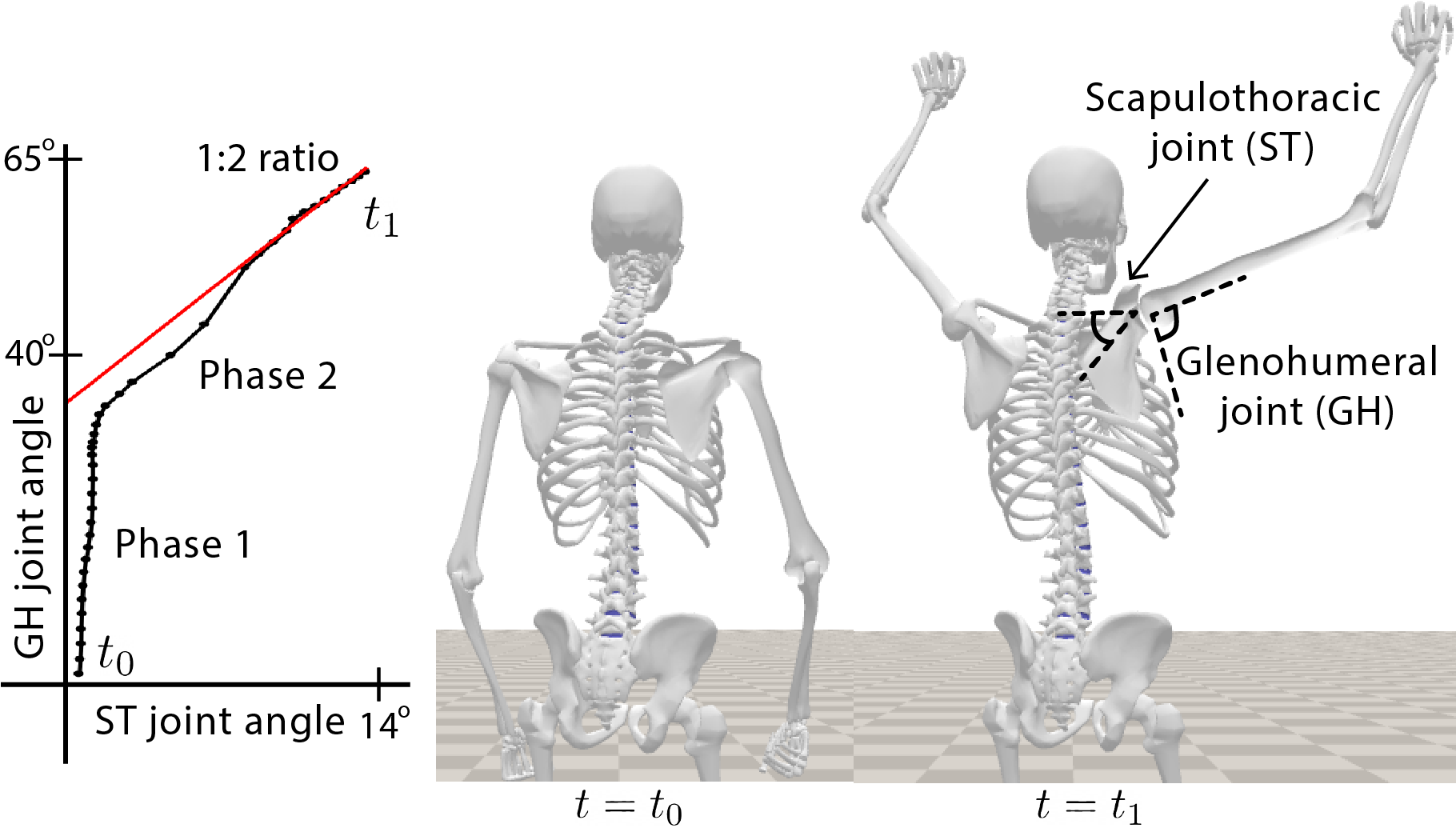}
    \caption{ Scapulohumeral rhythm. A unique phenomenon observed when humans rotate the upper arm latterally. The scapula elevation and humerus abduction maintains a fixed ratio after threshold configuration is reached. Our model (movement trajectory in black curve) is able to reproduce this phenomenon (shown in red curve).}
    \label{fig:scapula_rhythm}
\end{figure}

\subsubsection{Scapulohumeral Rhythm}

Scapulohumeral rhythm is one of the signature features observed from human shoulder movements \cite{rockwood2009shoulder}. It describes the relation between scapula elevation and humerus abduction. When a healthy subject laterally rotates the upper arm on the coronal plane, the glenohumeral joint (GH) and the scapulothoracic joint (ST) both contribute to the rotation (See Figure \ref{fig:scapula_rhythm}). The lateral rotation begins with the first phase called {\emph setting phase} where the rotation is almost purely done by the GH joint while the ST joint remains relatively unchanged. If the rotation continues beyond a threshold (around 30 degrees), both GH and ST joints start to rotate at a 1-to-2 ratio. This unique phenomenon emerges in our simulated motion from using sliding facet constraints between the scapula and the rib cage.

\subsubsection{Pressure Load on Disc}
\begin{figure}
    \centering
    \includegraphics[width=0.85\linewidth]{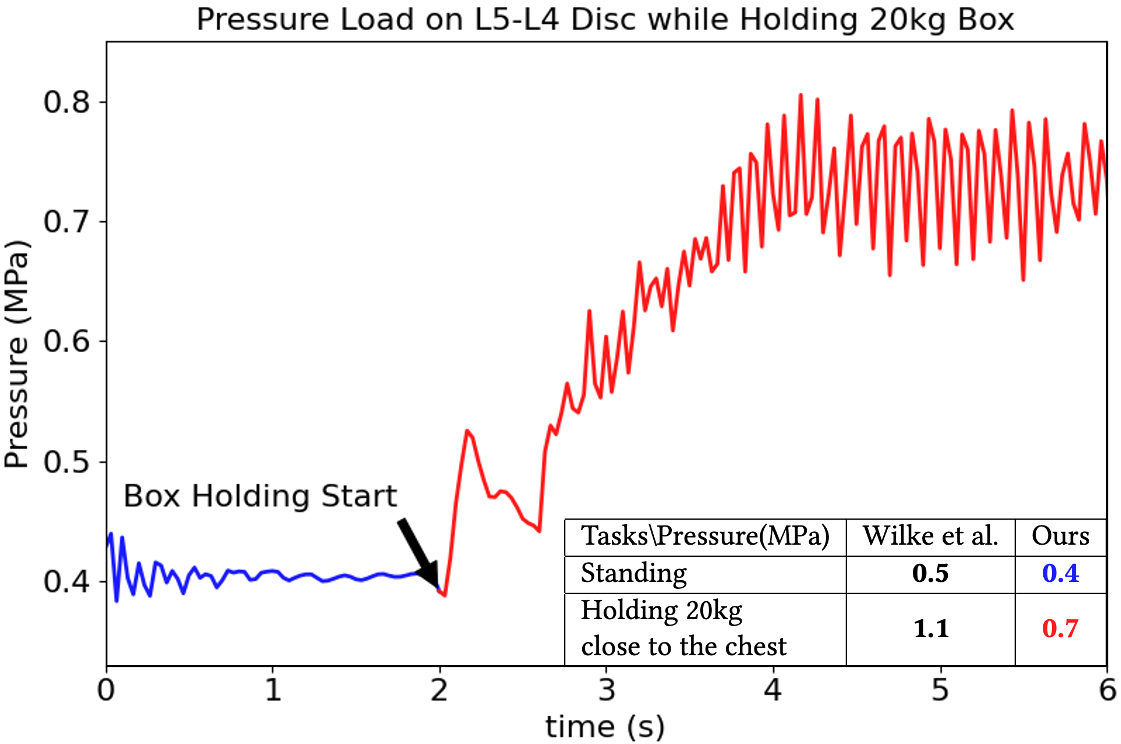}
    \caption{Pressure Load on L5-L4 Disc.}
    \label{fig:pressure}
    
\end{figure}
\textcolor{MyBlack}{
Existing biomechanics literature primarily focuses on spine and shoulder geometry and kinematics, such as lordosis (spine curvature), RoM, and Scapulohumeral Rhythm, while studies examining spine dynamics are less common due to the challenges of conducting in-vivo experiments. Among studies that measure dynamical properties, researchers often perform quasi-static analyses, such as assessing pressure loads on intervertebral discs. In \cite{wilke1999new}, the authors implanted a transducer with a pressure sensor into the L4-L5 disc, measuring loads during everyday tasks like standing, sitting, and holding objects. One experiment revealed a 0.5 MPa pressure during standing, which increased to 1.1 MPa when holding a 20 kg box and further to 1.8 MPa when holding the box 60 cm from the chest {(See Figure~\ref{fig:pressure}).}}

%

\textcolor{MyBlack}{To benchmark our model against their data, we compute pressure load by calculating elastic forces at the disc surface and dividing by the area attached to the bone (14.4 cm$^2$). Using a PD controller, we emulate a pose holding a box with feet welded to the ground, allowing the character to hold the box without learning or optimizing motion control. Our experiment demonstrates similar pressure for the standing task and an increase when holding a box, albeit less than the clinical data, possibly due to the PD controller's high gain. Holding a box should cause spinal bending and both rotational and translational deformation of discs, but the high gain results in negligible rotational disc deformation. Translational {DoFs} between bones are not controlled by PD (to prevent the spine from moving freely with magic forces) and are solely influenced by the translational disc deformation, which correctly increases during box carrying.}



\subsection{Ablations}

\textcolor{MyBlack}{In this section, we conduct ablation studies to demonstrate the roles of intervertebral discs and ligaments in our torso simulation model.}

\subsubsection{Disc} 
\textcolor{MyBlack}{The intervertebral discs play a critical role in preventing vertebral discs from colliding and rubbing against each other. When we remove the discs during motion retrofitting, we observe a significant increase in bone-to-bone collisions. To quantify the impact of the discs, we measure bone-to-bone displacements relative to rest-pose displacements and count the frames with collisions during retro-fitting, with and without the discs (See Figure~\ref{fig:bone_to_bone}). We approximate each vertebral bone as a cylinder to prevent the posterior side of bone (linked with the facets) from dominantly affecting distance measurements. We use mesh-to-cylinder distances to measure the displacements and compute the average and standard deviation (error bars) of the displacements across all motion frames in the dataset.}

\textcolor{MyBlack}{We observe that without the discs, the displacements at the lumbar bones are below 0 cm due to gravity, meaning that the distances between the bones are narrow. As a result, the bones collide with each other frequently at the incidence rate of 8.0\% and 14.1\% for L5 and L4 respectively during motion retrofitting. We also observe high collision rates among the upper thoracic bones, which are closer to each other at rest comparing to bones in other sections of the spine. With discs modeled (red in figure), collision rates are close to zero.}



\begin{figure}
    \centering
    \includegraphics[width=\linewidth]{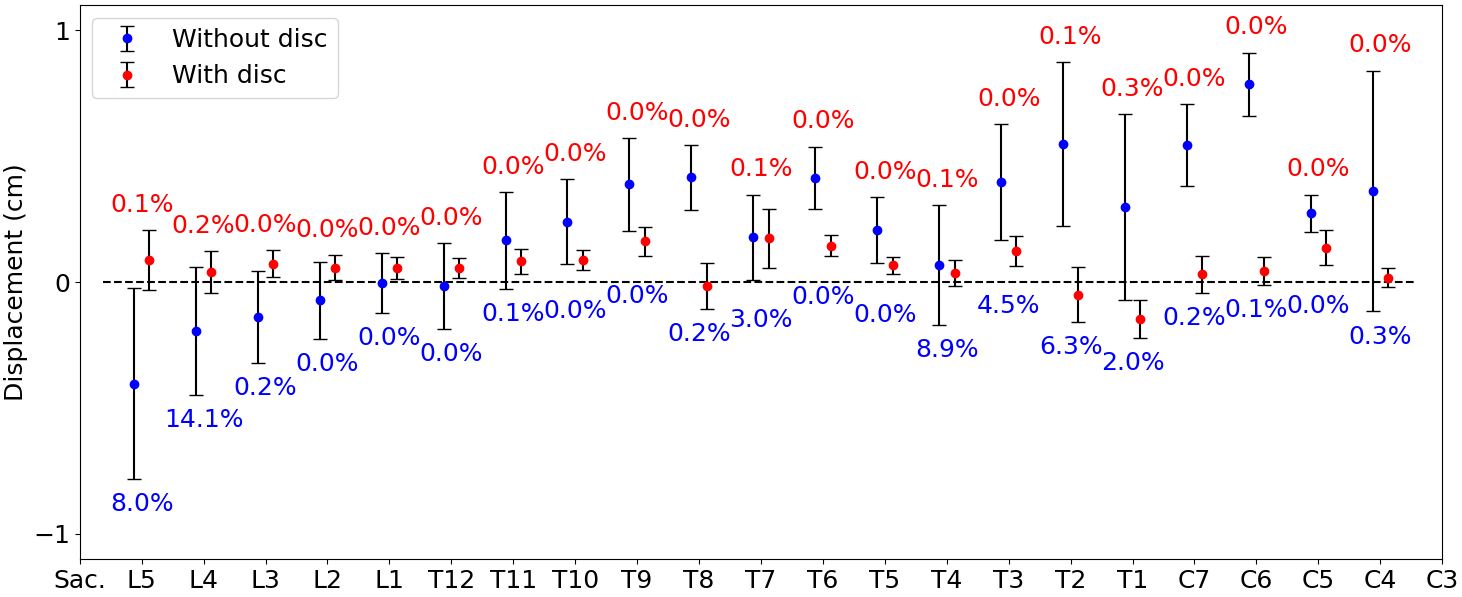}
    \caption{Bone-to-bone displacements and collision incidence rate during motion retrofitting.}
    \label{fig:bone_to_bone}
    
\end{figure}
\begin{figure}
    \centering
    \includegraphics[width=\linewidth]{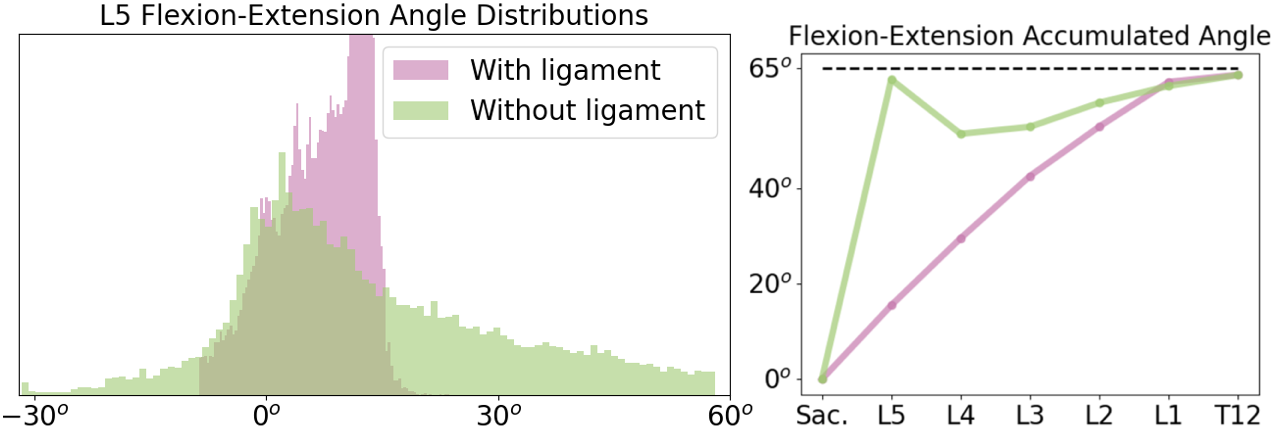}
    \caption{(left) We visualize distributions of flexion-extension angle of L5 during retrofitting the motions in the Mixamo dataset. (right) We accumulate the flexion-extension angles among lumbar bones during $65^o$ flexion.}
    \label{fig:flexion_distribution}
\end{figure}
\subsubsection{Ligament} 

\textcolor{MyBlack}{
In addition, we also study the impact of ligaments. Removing the ligaments results in a unrealistically wider range of motion (RoM) than that of a real human spine (See Figure~\ref{fig:flexion_distribution}). Without the ligaments, the L5 exhibits a large flexion-extension angle ranging from $-30^o$ to $60^o$. When retrofitting the spine to the existing motions, our method tends to find extreme solution where only L5 rotates to fit the motion while other vertebral bones remain in the rest pose (See Figure~\ref{fig:flexion_distribution} Right). This experiment shows that the ligaments with identified parameters provide accurate RoM,  significantly improving our simulation model in producing more human-like results.}
\section{Discussion}

This paper introduces a novel human torso model featuring key anatomical elements from the spine and the shoulder complex. We show that our model is robust enough to retrofit a large and diverse motion database, and to be readily incorporated into existing full-body skeletons to simulate highly dynamic movements. We further partially validate that the statistics of our model is consistent with the human data from biomechanics literature. While there has been some attempts from the biomechanics community to build similarly detailed spine or shoulder models, we demonstrate in this work that tools, algorithms, and implementation know-hows that continue to mature in Computer Graphics can make anatomically realistic human models robust and efficient for computationally-heavy applications, accelerating wide adoption of these models.


Our current implementation has a few limitations. First, we have yet to model muscles on the torso which drive the spine and the shoulder. Our detailed skeletal model should help the development of an accurate torso muscle model. Second, passive dynamics of the joints and the discs, such as spring-damping coefficients and Poisson ratio, are tuned manually. While measurement of dynamic properties of human skeleton is much less available compared with kinematics, our future work would include matching passive dynamics of our model to experiment data. Third, our current retrofitting is solved frame-by-frame and can lack temporal smoothness in some cases. Optimizing over a time window, or better initialization scheme would improve the quality of retrofitting.

Perhaps the most exciting application domain enabled by our model lies in medicine and healthcare. \textcolor{MyBlack}{
The proposed computer graphics techniques have the potential to contribute to the development of noninvasive diagnostic tools for analyzing and predicting internal spinal conditions during daily activities. For instance, back pain, a prevalent medical issue affecting 80\% of individuals at some point in their lives, is often caused by slipped (herniated) discs that impinge on the spinal nerve cord. The depth of this impingement affects the severity of pain and varies with different poses. Our model could simulate such protrusions and anticipate how various full-body poses during a gait cycle influence the depth of impingement. Furthermore, incorrect squatting posture during strength training, leading to an improperly bent spine, can increase pressure on the spinal column. Our method may be capable of reproducing these phenomena and predicting the optimal squat position to minimize spinal load. Considering the wide variability in torso {RoM} due to factors like age, sex, and physical ability, personalizing our model for medical applications presents another promising future direction.}

\textcolor{MyBlack}{
Our framework also holds promise for enhancing computer animation applications. Existing skinning techniques such as linear blend skinning and dual quaternion skinning can be readily integrate to our model. In addition, by accounting for detailed bone meshes, our spine model can enable novel skinning techniques to visualize the outlines of scapular and vertebral protrusions commonly seen in actual human skin. Our model can thus serve as the foundation for a novel skinning pipeline. Additionally, adapting our spine model to quadruped animals, such as cats, may lead to improvements in the realism of animated quadruped motion.
}

\begin{acks}
This work was supported by Wu Tsai Human Performance Alliance at Stanford University, the Stanford Institute for Human-Centered AI (HAI), and NSF:FRR 2153854.
\end{acks}
\bibliographystyle{ACM-Reference-Format}
\bibliography{reference}

\end{document}